\documentclass[sigconf]{acmart}

\usepackage{hyperref}
\usepackage{graphicx}
\usepackage{booktabs} 
\usepackage{bm}
\usepackage{setspace}
\usepackage{epstopdf}
\usepackage{float}
\usepackage{subfigure}
\usepackage{natbib}
\usepackage{enumitem}

\graphicspath{{F:/OriginPro8/cx/social_network_evolution/all_graph/}}

\setcopyright{rightsretained}

\begin{document}

\title{Modeling Personalized Dynamics of Social Network and Opinion at Individual Level}

\author{Xi Chen}
\affiliation{%
  \institution{Central South University}
  \city{Chang Sha}
  \state{China}
}
\email{xchen@csu.edu.cn}

\author{Jie Tang}
\affiliation{%
  \institution{Tsinghua University}
  \city{Beijing}
  \state{China}
}
\email{jietang@tsinghua.edu.cn}

\author{Yizhou Sun}
\affiliation{%
  \institution{University of California, Los Angeles}
  \city{Los Angeles}
  \state{CA}
}
\email{yzsun@cs.ucla.edu}

\renewcommand{\shortauthors}{X. Chen et al.}


\begin{abstract}
Network dynamics has always been a meaningful topic deserving exploration in the realm of academy. previous network models contain two parts: (1) generating structure as per user property; (2) changing property as per network structure. Properties in these models, however, cannot be interpreted to concept in prevalent social theories or empirical truth. Also, they usually treat everyone in an uniform fashion. While such assumption is quite misguiding, and thus saliently limits their performance. To overcome these flaws, we devise a personalized evolving model for social network and opinion (PENO), where citizens' ideology is revealed by variable opinions and four dimensions of personality are considered for each entity - leadership, openness, agreeableness, and neuroticism. Opinion propagates via social tie, tie generates from opinion affinity, and personalities integrally work with opinion and tie across evolution. To our best knowledge, PENO is the first attempt to introduce personality impact in network dynamics and verify social science with reasonable visualization during simulation. We also present its probabilistic graph and conceive iterative learning algorithm. Experiments show PENO outperforms several state-of-the-art baselines over two typical prediction tasks - congress voting prediction for legislative bills and friendship prediction on a book-commenting website. Finally, we discuss its scalability to do multi-task learning and transfer learning in daily scenarios.
\end{abstract}

%
%


\ccsdesc[500]{Social network~Information network}
\ccsdesc[300]{Dynamic network~Network evolution}
\ccsdesc[300]{Social relationship~Tie}

\keywords{social network, dynamic network, personality, human-centered application}

\maketitle

\section{Introduction}
As the variety and popularity of information networks increase, network study have become a prevalent topic in academia and industry. Understanding how network evolves over time, not only help us to mine potential partners in future, but also provides advice to plan travel route, predict election outcomes and a variety of social applications. What is more, it serves as a guideline to regulate group, which is crucial for managers like government and enterprises. However, network evolution involves both temporal association and spatial interaction, making it an incredibly complex process for research: In the one hand, individuals form ties or break old ones as per their property (e.g. idealogy or interest); In another hand, everyone convey his unique impact to friends and receive reciprocally, thus properties propagate via social ties over time.

Due to the significance of understanding evolution, many works have been conducted to analyze and model network dynamics\cite{Barabasi1999Emergence, Leskovec2005Realistic, Leskovec2008Microscopic, Gruhl2004Information, Sarkar2005Dynamic, Heaukulani2013Dynamic}. But most factors in these works are not explainable to fit recognized social theories or empirical truth. And they usually treat everyone without any distinctiveness. In fact, individual presents innate diversity in social activities\cite{Inbar2012Political}. This misleading assumption strongly limits their performance. Inspired by the concept of information propagation \cite{Chen2013Information} and widely recognized psychology foundation\cite{MURRAY2010THE, Northouse2014Leadership}, we propose a personalized evolving model for social network and opinion (PENO) to resolve these issues: Variable properties represent citizens' opinion and four kinds of personality factors are considered for each individual - leadership, openness, agreeableness, and neuroticism. Opinion propagates via social tie, tie generates from opinion affinity\cite{Mcpherson2001Birds}, and personalities determine the willing or capacity how greatly a person influence others (leadership), receive other views (agreeableness), make novel friends (openness), and be disturbed by exception from environment (neuroticism).
\begin{figure}[tt]
\setlength{\belowdisplayskip}{5pt}
        \includegraphics[width=9cm,height=6cm]{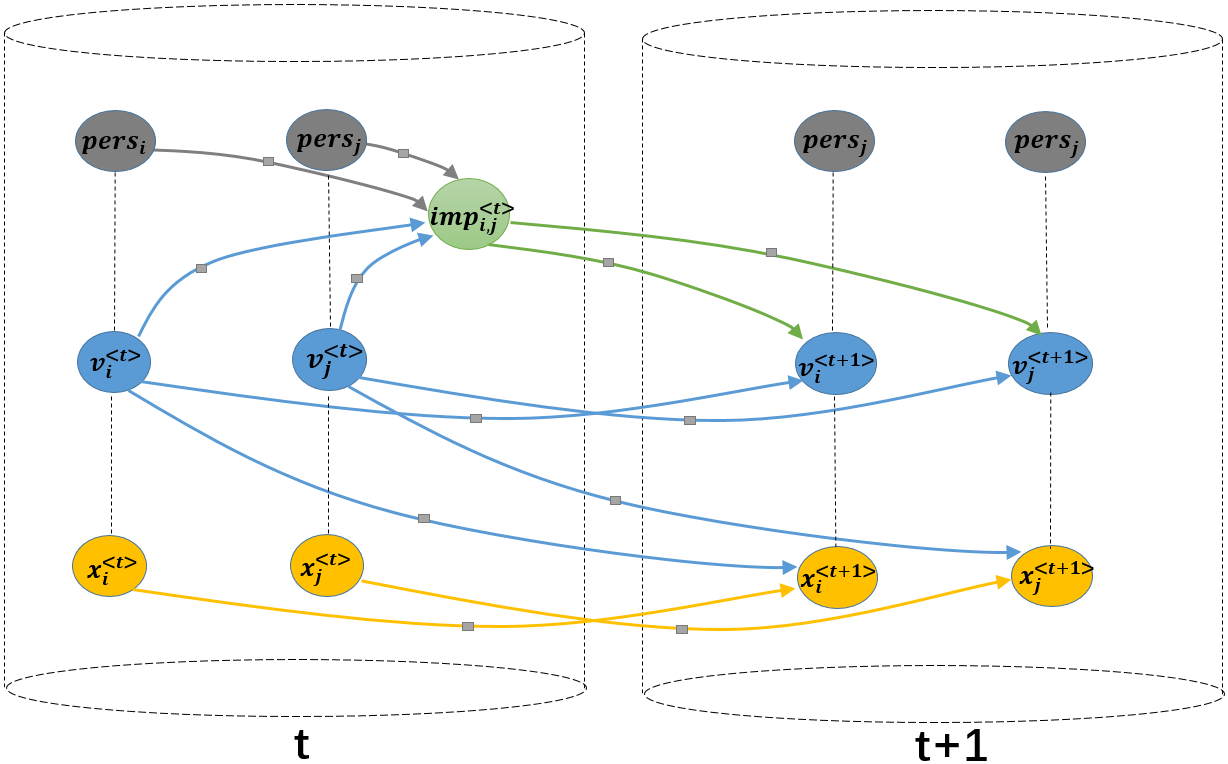}
        \caption{opinion propagation from moment $t$ to moment $t+1$ (yellow nodes are opinion, blue nodes are opinion trends, and grey nodes are personalities,green node is mutual social impact between persons)}
\end{figure}

Saliently distinguished from previous works, we aim to understand the dynamics of social network from idealogy and psychology perspective. To this end, we investigate opinion propagation mechanism and the feature of opinion for forming ties at individual level. A novel iteration model is carefully devised where network topology and personal idealogy evolves together. Simulation shows that our model is promising in verifying social science (e.g. opinion convergence\cite{Sobkowicz2011Opinion} and leadership effect\cite{Northouse2014Leadership}). By learning opinions and personalities from observed data, our model also significantly outperforms the state-of-art dynamic network approaches in two typical tasks -
predicting legislative votes and the friendship on an article-sharing website. The distribution of learnt personalities also reveal group feature and discover some actual leaders over periods. Its potential in human-centered applications with respect to multi-task learning and transfer learning is discussed.

To our best knowledge, this is the very first attempt to incorporate personality effect into dynamic network model. The contribution in this paper includes:
\begin{itemize}[leftmargin=8pt]
\item We propose a unified model that captures correlated evolution feature of network topology and latent properties at individual level. Thanks to carefully devised mechanism, our model can be visualized during simulation. This provides a promising method to verify social science through computation.
\item Given network graph series data, our model is capable to learn individual properties from observed history. Experiments not only show our advantage over the state-of-art approaches in two typical prediction tasks, but also reveal social feature of a group and discover some of its actual leaders.
\item We give the first attempt to model network dynamics from idealogy and psychology perspective by introducing personality. In fact, it serves as an innate property for entity so that can be shared in a variety of human-centered applications. Our model implies a novel vision to ground multi-task learning or transfer learning on network evolution.
\end{itemize}
\section{Preliminary Description}
In dynamic social network, individuals can be expressed as a group of nodes. For a pair of nodes, whether they are tied with the other indicates their mutual relationship. Since social behavior is usually reciprocal, global relationship variation can be described as a time serial of undirected network graph
\emph{$F = \{G^{<0>},G^{<1>},...\}$}, where \emph{$G^{<t>}$} consists of all ties at moment $t$. Each individual possess a property vectors namely "opinion". Opinion represents his or her ideology stand and varies with time as well. Thus all opinions compose another time serial \emph{$U = \{X^{<0>},X^{<1>},...\}$}. Besides moving opinion, nodes are also associated with their respective personalities $\zeta=[R,L, \sigma,B]$, determining their social patterns at individual level during network evolving.
\section{Network Model}
Opinion and social tie are two key elements interacting with each other during evolution process. On the one hand, individuals in the group pay much attention to friends' behavior and claims. According to "social influence" theory \cite{Crandall2008Feedback, Kempe2000Maximizing, Tang2009Social}, they will more or less agree with others over time, and thus lead to the convergence of group opinion \cite{Giddings1970The}. On the other hand, people have natural inclination to seek for cognitive affinity, so tie is usually formed or kept between those who hold similar views \cite{Mcpherson2001Birds}.

Besides common pattern, a certain individual is also driven by innate personality and presents his own social fashion. For example, whether someone is extroversive or introversive can largely determine his will in friend making; while tolerance and agreeableness helps him or her to accept other views, etc.

Putting personality-based opinion propagation and ties formation together, we then propose the personalized evolution model for social network and opinion (PENO), where evolution iterates in an succinct way: (1) propagating opinion in light of social influence; (2) building social ties as per opinion affinity. In rest of this section, we will introduce model details and show some interesting phenomenon in simulation.
\begin{figure*}[tt]
        \begin{minipage}{5.5cm}
            \setlength{\belowcaptionskip}{-0cm}
            \includegraphics[width=5.5cm]{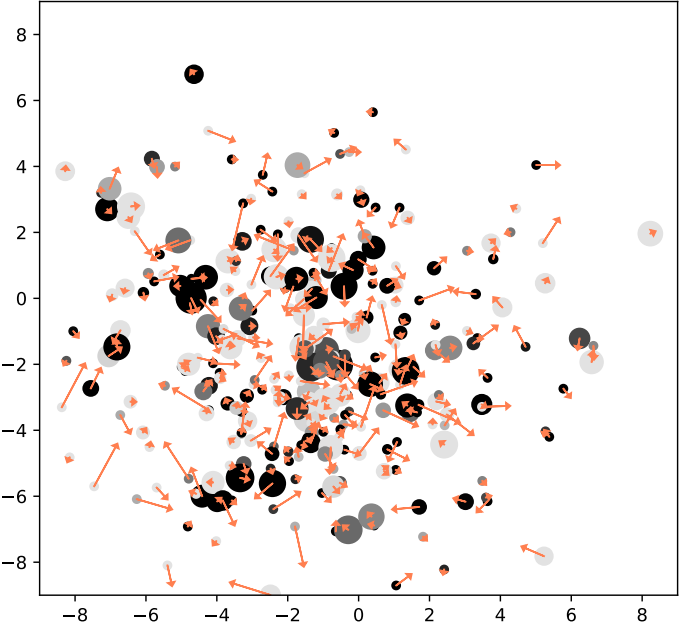}
            \caption* {(a) moment = 0}
        \end{minipage}
        \begin{minipage}{5.5cm}
            \setlength{\belowcaptionskip}{-0cm}
            \includegraphics[width=5.5cm]{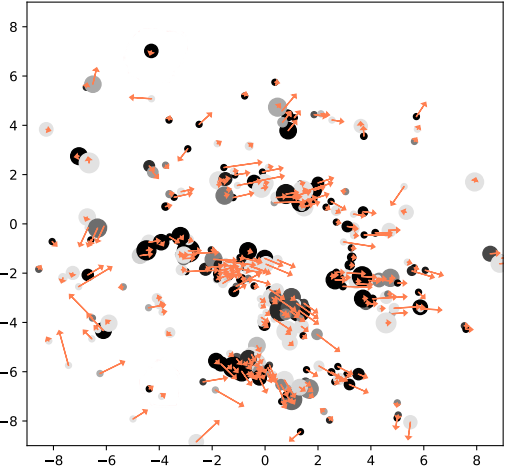}
            \caption* {(b) moment = 80}
        \end{minipage}
        \begin{minipage}{6.6cm}
            \setlength{\belowcaptionskip}{-0cm}
            \includegraphics[width=6.5cm]{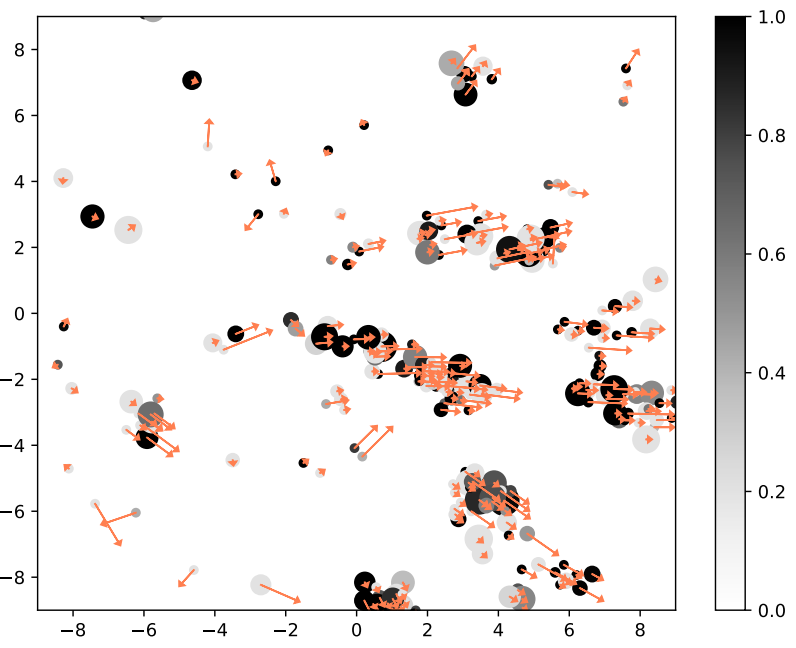}
            \caption* {(c) moment = 200}
        \end{minipage}
    \caption {Entities migration in social network.}
\end{figure*}
\begin{figure*}
    \begin{minipage}{5.5cm}
        \setlength{\belowcaptionskip}{-0cm}
        \includegraphics[width=5.5cm]{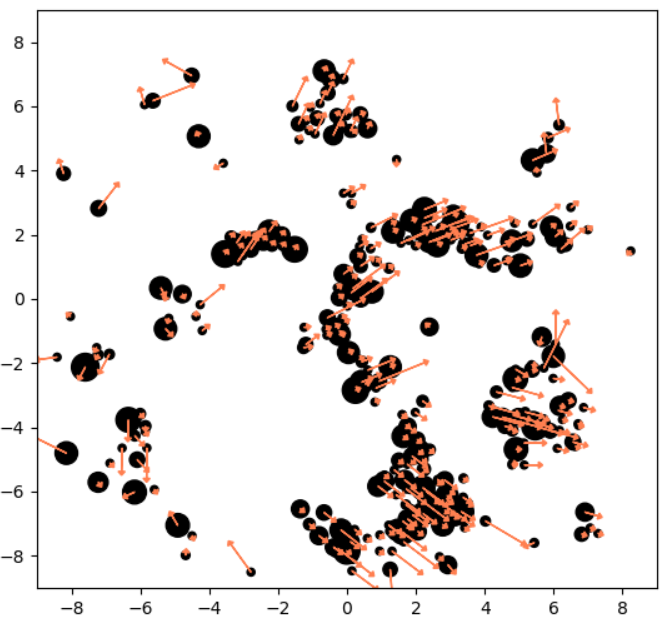}
        \caption*{(a)$r \sim N(1,1)$}
    \end{minipage}
    \begin{minipage}{5.5cm}
        \setlength{\belowcaptionskip}{-0cm}
        \includegraphics[width=5.5cm]{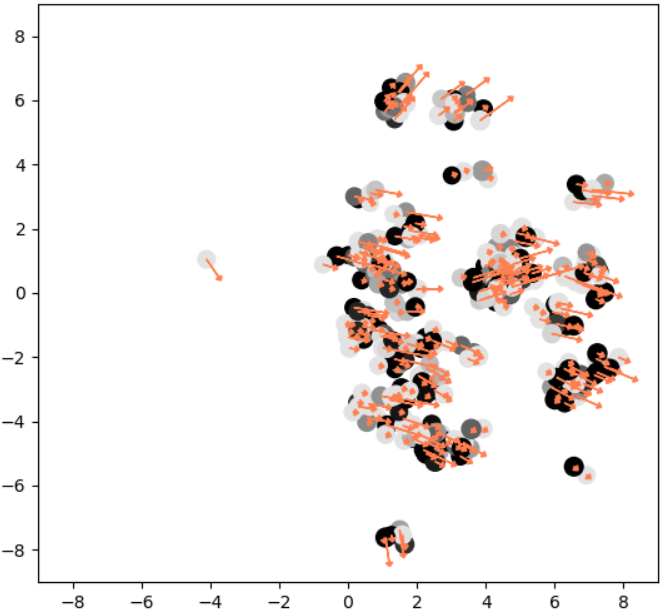}
        \caption*{(b)$b \sim N(3,1)$}
    \end{minipage}
    \begin{minipage}{6.6cm}
        \setlength{\belowcaptionskip}{-0cm}
        \includegraphics[width=6.5cm]{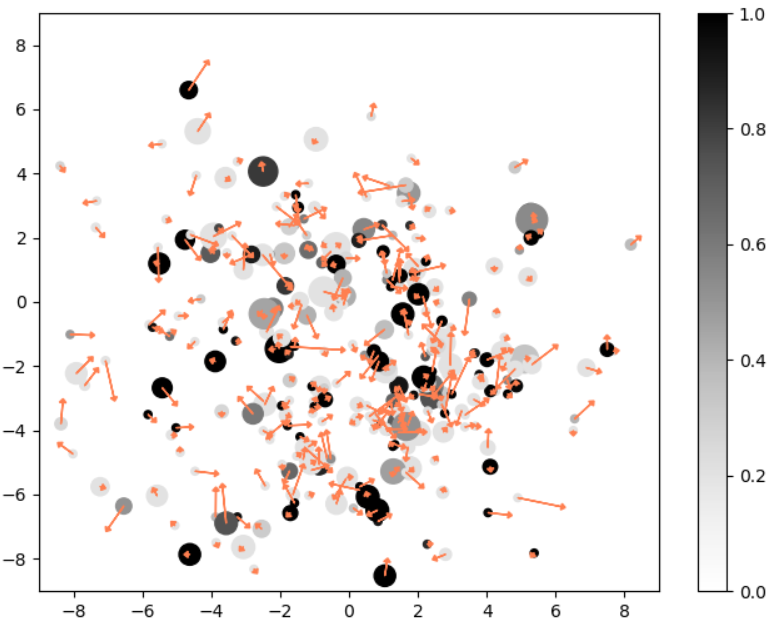}
        \caption*{(c)$\sigma \sim N(4,1)$}
    \end{minipage}
    \caption {Migration of entities with different personalities (moment = 200).}
\end{figure*}
\begin{figure*}[tt]
\begin{minipage}{5.5cm}  
        \setlength{\belowcaptionskip}{-0cm}
        \includegraphics[width=5.5cm]{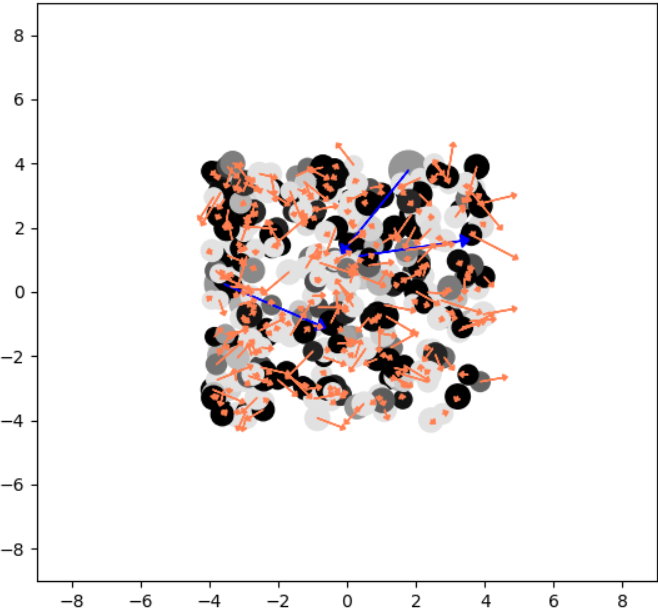}
        \caption*{(a)states with volatile leaders at initial moment.}
\end{minipage}
\ \ \ \
\begin{minipage}{5.5cm}
        \setlength{\belowcaptionskip}{-0cm}
        \includegraphics[width=5.5cm]{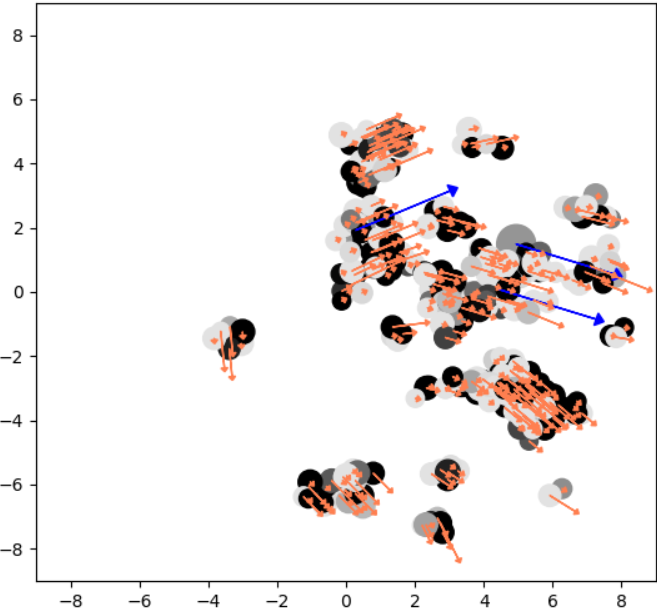}
        \caption*{(b)states with volatile leaders after moments.}
\end{minipage}

\begin{minipage}{5.5cm}
        \setlength{\belowcaptionskip}{-0cm}
        \includegraphics[width=5.5cm]{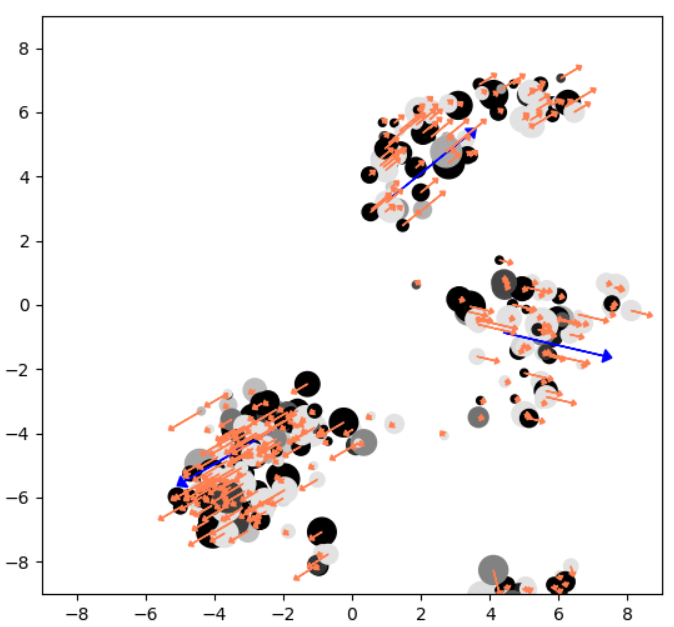}
        \caption*{(c)states with steady leaders after moments.}
\end{minipage}
\ \ \ \
\begin{minipage}{5.5cm}
        \setlength{\belowcaptionskip}{-0cm}
        \includegraphics[width=5.5cm]{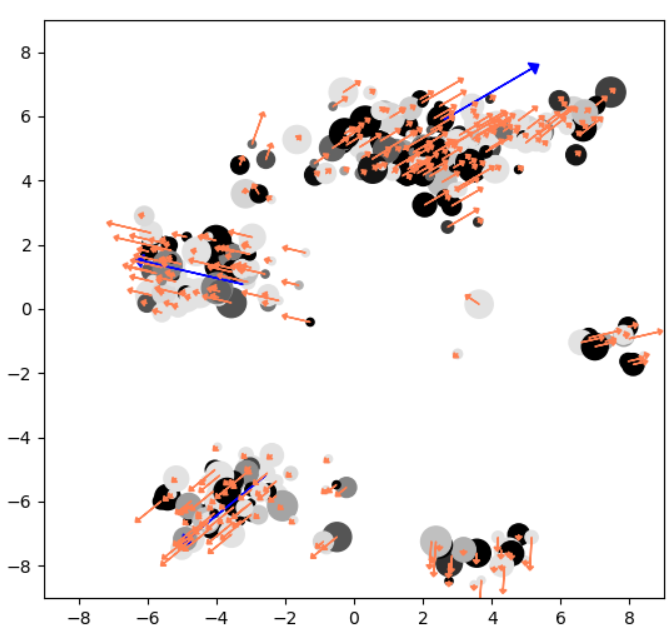}
        \caption*{(d)states with steady leaders (inverse trend).}
\end{minipage}
\caption {Entity migration with different kinds of leaders (l = 20).}
\end{figure*}
\subsection{Opinion Propagation}
To make PENO model clear, at first we have to decide the mechanism how individual opinion propagates via social ties. As discussed above, opinion is expressed as property vector in our proposition. Thus modeling evolution implies to depict the association between current opinion vector and those in the past. Since network evolution will become an incredibly complex process if extracting overmuch features as to time and topology, here we only take adjacent moments into consideration (e.g. $x^{<t>}\rightarrow x^{<t+1>}$).

Previous work models property change very straightforwardly: forcing properties in adjacent moments to be similar via a variety of regulation/prior (e.g. Gaussian prior) \cite{Corneli2016Modelling, Sarkar2005Dynamic}. Admittedly it makes property vary in a smooth fashion without abrupt mutation. While it is also likely to cause meaningless vibration and refer none of topology information. Then another idea came out to overcome its flaw: For each entity, weighting and combining his current property and neighbors' property as the property in next moment. For instance \cite{Heaukulani2013Dynamic}:
    {\setlength\abovedisplayshortskip{1pt}  \setlength\belowdisplayshortskip{1pt}
    \begin{equation}
    \setlength{\abovedisplayskip}{3pt}
    \setlength{\belowdisplayskip}{3pt}
        x_n^{<t+1>} \sim N( \overline{x_n^{<t>}},{\sigma ^2})
    \end{equation}
where $\overline{x_n^{<t>}}$ is the property average of person $n$ and all his neighbors.
Such genre includes social impact compared to the earlier. But it turns out nodes will easily be trapped into communities and stop moving in this case \cite{Sarkar2005Dynamic, DBLP_conf_kdd_GuSG17}.

In a recent study, the concept of velocity are introduced to relieve the "community trap" phenomenon \cite{DBLP_conf_kdd_GuSG17}. In nature, velocity represents the moving trend of position. Likewise, node velocity directly draws property variation between adjacent moments. Both statics and abrupt change can be avoided as long as velocity is kept non-zero and slight. This analogy also have intuitive explanation to opinion dynamics in social network: opinion stands for one's ideology and thus gradually grows with living experience over time. When someone spends time on personal affairs, his opinion migrates along his previous track rather than staying. At some other time, if people are taking social activity in a group, though their opinion will not be immediately affected by each other, opinion tracks tend to be similar seeking for "affinity".

Based on above description, opinion should vary as per the following formula:
    \begin{equation}
    \setlength{\abovedisplayskip}{3pt}
    \setlength{\belowdisplayskip}{3pt}
        \vec{x}_n^{<t+1>} = \vec{x}_n^{<t>} + \vec{v}_n^{<t>}
    \end{equation}
where $x$ refers to opinion. To simplify computation, velocity value is fixed in rest of the paper, so that we only need to consider its direction. Vector size of opinion and direction are assumed to be 2 dimension:
    \begin{equation}
    \setlength{\abovedisplayskip}{3pt}
    \setlength{\belowdisplayskip}{3pt}
        \vec{x}_n^{<t>} = ({z}_n^{<t>},{c}_n^{<t>})
    \end{equation}
and
    \begin{equation}
    \setlength{\abovedisplayskip}{3pt}
    \setlength{\belowdisplayskip}{3pt}
        \vec{v}_n^{<t>} = {v}_n^{<t>} \cdot (cos\theta_n^{<t>}, cos\theta_n^{<t>}),
    \end{equation}
which helps to visualize entities in simulation. Surely it can be extended to any positive integer and do no harm to model scalability.

\textit{{\bfseries personality.}} \noindent
Captured common feature of opinion dynamics, we draw a outline of propagation that treat everyone uniformly. While social group consists of various kinds of individuals and performs locally. For instance, politicians prefer to lead a team and devote into persuading others; while some others are good at listening ideas and accept new views very soon, referred as "follower" in study \cite{Chaminade2002Leader}. Here we formally introduce a core concept - personality. By setting several major personality genre as independent parameters for each person, PENO model owns much more power to grasp social behavior at individual level than those discussed above. Fusing Big Five Character theory and concepts in cutting-edge study \cite{Staiano2012Friends, Kosinski2014Manifestations, Bachrach2012Personality,Zhao2017WHO}, we eventually introduce four kinds of personality: openness, leadership, and neuroticism, agreeableness. As per their working fields, three kinds are interpreted below, and agreeableness will be referred in tie formation section. Since social impact are regarded as updating the direction of opinion velocity, overall impact to person $i$ should be the weighted sum of impact from his neighbors:
    \begin{equation}
    \setlength{\abovedisplayskip}{3pt}
    \setlength{\belowdisplayskip}{3pt}
        imp_i^{<t>} = \mathop{\sum}\limits_{(i,k) \in G^{<t>}}\frac{l_k}{\lambda_i^{<t>}}\triangle\theta_{ki}^{<t>}
    \end{equation}
where $l_k$ indicates leadership. leadership shows the power of a person persuading friends to be alike himself, i.e. with similar opinion trend. $\lambda_i^{<t>}$ is the leadership sum of person $i$'s friends. It serves as a normalization factor to adaptively restrict the range of new opinion trend $\theta_i^{<t+1>}$. Thus value for leadership can be any positive number. $\triangle\theta_{ki}^{<t>} = \theta_{k}^{<t>} - \theta_{i}^{<t>}$ reflects trend difference between opinion of person $k$ and $i$.
Through propagation, opinion trend mixes with social impact it receives and thus form a new trend, expressed as
    \begin{equation}
    \setlength{\abovedisplayskip}{3pt}
    \setlength{\belowdisplayskip}{3pt}
        \theta_i^{<t+1>} \sim N(\theta_i^{<t>}+ r_{i}\times imp_i^{<t>}, \sigma_i^2)
    \end{equation}
$r_i$ is agreeableness and $\sigma_i$ is neuroticism: agreeableness $r_{i}$ determines the degree a person conform to others' opinion. The bigger $r_{i}$, the more acceptance. In boundary cases, if $r_{i} = 0$, person $i$ will not believe any heard claims. His opinion varies singly. Otherwise when $r_{i} = 1$, the person totally follow friends regardless of himself. Neuroticism $\sigma_i$ measures how greatly opinion deviates the expected track, considering noise and exception in environment. Its value should not be less than 0.

Figure 1 presents exactly how our opinion propagation works between a pair of friends $i$ and $j$ at certain moment $t$. As shown, social impact is determined by personalities and opinion trends, then the impact and their current trends contribute to future trends at next moment $t$. And opinion of each person migrates as per his current trend.
\subsection{Tie Formation}
Similar to propagating opinion, collective feature of individuals also play an important role in determining social tie, as described in this section.

Social tie reveals contact information of network. Such information, however, is not static over time. Individual continuously make new friends through a variety of daily activities. But there is no way someone can keep in touch with all former friends: limited energy and time for consuming. Furthermore, friendship also ends when two no longer share common views. To accommodate this phenomenon, network does not inherit existed ties till new moment. Instead, they will be formed at each moment in our model.

Recalling latent space model\cite{Hoff2002Latent}, we should devise an evaluation function to score all pairs of persons, and then the score determine corresponding tie will be formed or not. Spontaneously we can take the advantage of introduced properties i.e. opinion and personality to make evaluation.

As per description about affinity, individuals appreciate neighbors who hold alike views and prefer to contact them. Hence in fact the function aims to evaluate opinion similarity. Assuming each dimension of opinion have equal importance, Euclidean distance is selected as the measure of similarity. With Consideration of model scalability, we further adapt Gaussian function to form ties. Specifically, when we expect to determine tie between person $i$ and $j$ at moment $t$, evaluation function is mapped to a probability
    \begin{equation}
    \setlength{\abovedisplayskip}{3pt}
    \setlength{\belowdisplayskip}{3pt}
        p_{ij}^{<t>} = exp(\frac{-1}{\xi^{2}}\frac{||x_i^{<t>} - x_j^{<t>}||^{2}}{b_{i}b_{j}})
    \end{equation}
and
    \begin{equation}
    \setlength{\abovedisplayskip}{3pt}
    \setlength{\belowdisplayskip}{3pt}
        e_{ij}^{<t>} \sim Bernoulli(p_{ij}^{<t>})
    \end{equation}
where $\xi$ is a system parameter to control the overall tie formation difficulty. openness $b_n$ draws ones' willing in making friends. Note that a pair ($b_i$,$b_i$) multiplies in the formula so that they have to work together, indicating friends need mutual support. Obviously $b_n$ also cannot be negative. Tie will be generated in the probability $p_{ij}$.
\subsection{Framework}
Integrate opinion propagation and tie formation into an iterative process, network evolution become complete in our model. At each moment $t$, network topology is formed given opinion and openness of each entity (affinity), then opinion trends are updated according to the topology and other kinds of personality(social influence), determining opinion at next moment $t+1$. Entire evolution process can be described as Algorithm 1 presents.
\section{Simulation}
To validate the effectiveness of our model and study dynamic feature of group, we conduct simulation experiment and visualize some interesting results.

\textit{{\bfseries Settings.}}Here we provide evolution as to 300 entities in a variety kinds of settings. For initialization, each nodes are assigned a two-dimensional coordinate vector as per normal distribution, whose mean is 4. Visualization scale is set to four times of the mean i.e. 16 for birds eye observation. Opinion velocity is 0.03, minute enough to smooth evolution process. These settings are identical across simulation section. Since how personalities contribute to social connection is the most interesting topic in our model, we mainly investigate their impact on evolution, as shown below.

Figure 2 presents the benchmark so that results in other figures are compared with it for analysis. Basic settings for personalities are $r \sim N$(0.6,1), $l \sim N(0.3,1)$, $\sigma \sim N(1,1)$, $b \sim N(0.3, 1)$. They are also subjected to range constraints as discussed in model section. Other figures are with the same if not specifically mentioned. In figures, node placement refers to opinion, arrow is trend. Node color, node size, and arrow length corresponds to openness, agreeableness, and leadership respectively.

\begin{figure*}[tt]
    \includegraphics[width=12cm]{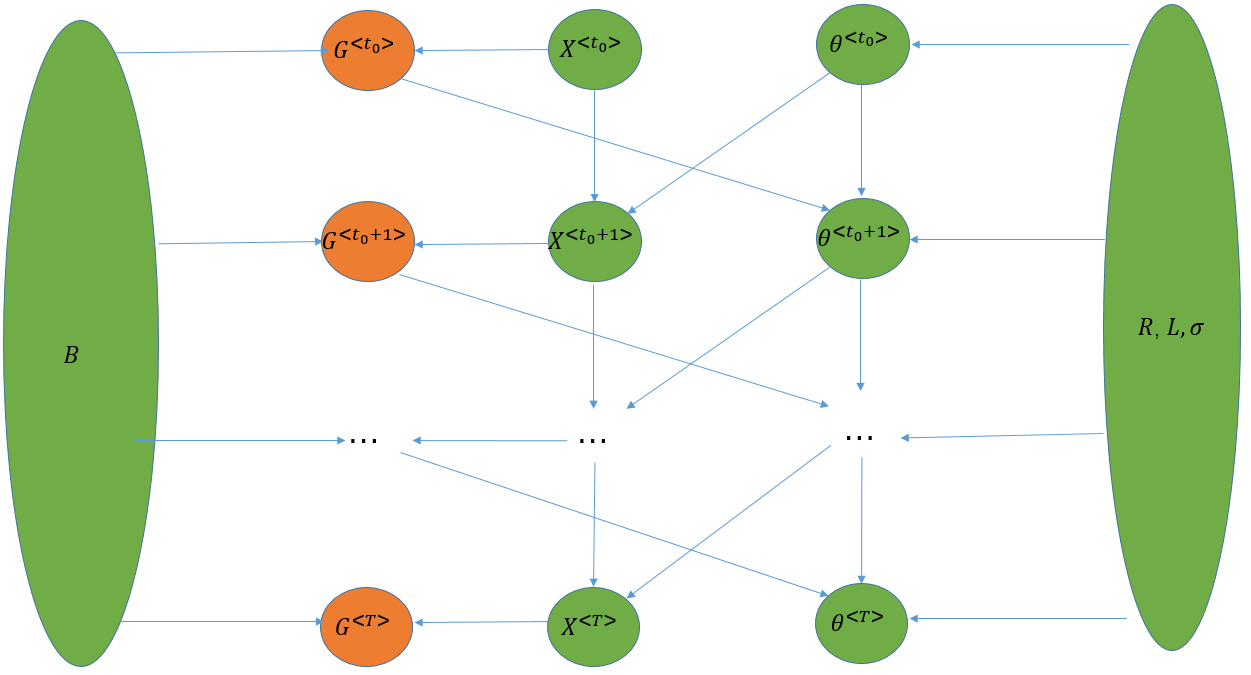}
    \caption{Probabilistic graph expression of PENO model.}
\end{figure*}

\textit{{\bfseries Phenomenons.}}
Thanks to these visualization skills, we observe some intuitive phenomenons from Figure 2:\par
(1) Outlier moves merely along his own direction. As in daily scenarios, if someone holds a distinct view, communication becomes much harder than usually, so does making friends. Inversely, limited guidance from others will lead to more deviation to common idea. Therefore, maybe he never take part in social interaction at all.\par
(2) Entity with high openness (black) are much likely to follow the mainstream opinion. This is also reasonable: mainstream are formed by opinions of many entities. Openness indicates one's willing to accept other opinion. In this case, his original opinion cannot be compared with mainstream over time. \par
\begin{tabular}[t]{l}
    \toprule
    {\bfseries Algorithm 1: Personalized evolution model}\\ \ \ \ \ \ \ \ \ \ \ \ \ \ \ \ \ \ \ \ \
    {\bfseries for social network and opinion} \\
    \midrule
    {\bfseries 1:} {\bfseries For} t = 0 to T: \\
    {\bfseries 2:} \ \ \ {\bfseries If t == 0}: \\
    {\bfseries 3:} \ \ \ \ \ \ // initialization \\
    {\bfseries 4:} \ \ \ \ \ \ {\bfseries For} n = 1 to N: \\
    {\bfseries 5:} \ \ \ \ \ \ \ \ \ initialize personalities $r_n,l_n,\sigma_n,b_n$ \\
    {\bfseries 6:} \ \ \ \ \ \ \ \ \ initialize opinion $x_n^{<0>}$ and opinion trend $\theta_n^{<0>}$ \\
    {\bfseries 7:} \ \ \ \ \ \ {\bfseries End} \\
    {\bfseries 8:} \ \ \ {\bfseries Else}: \\
    {\bfseries 9:} \ \ \ \ \ \ // opinion propagation \\
    {\bfseries 10:} \ \ \ \ \ {\bfseries For} n = 1 to N: \\
    {\bfseries 11:} \ \ \ \ \ \ \ \ calculate social impact $imp_n^{<t>}$ \\
    {\bfseries 12:} \ \ \ \ \ \ \ \ sample trend $\theta_n^{<t>}$ given $\theta_n^{<t-1>}$,$imp_n^{<t>}$,$r_n$,$\sigma_n$\\
    {\bfseries 13:} \ \ \ \ \ \ \ \ update opinion $x_n^{<t>}$ given $x_n^{<t-1>}$, $\theta_n^{<t-1>}$\\
    {\bfseries 14:} \ \ \ \ \ {\bfseries End} \\
    {\bfseries 15:} \ \ {\bfseries End}: \\
    {\bfseries 16:} \ \ // tie generation \\
    {\bfseries 17:} \ \ {\bfseries For} i,j = 1 to N: \\
    {\bfseries 18:} \ \ \ \ \ calculate probability $p_{ij}^{<t>}$ given $x_i^{<t>}$,$x_j^{<t>}$,$r_i$,$r_j$ \\
    {\bfseries 19:} \ \ \ \ \ generate tie $e_{ij}^{<t>}$ in the probability $p_{ij}^{<t>}$ \\
    {\bfseries 20:} \ \ {\bfseries End} \\
    {\bfseries 21:} {\bfseries End} \\
  \bottomrule
\end{tabular}\\ \\ \\
(3) Opinion gathers into different communities and continuously attracts new members. Take parties as example, republicans exactly have identical vision and absorb individuals who are devout to it.

\textit{{\bfseries Analysis.}}
Figure 2(a)-(c) have showed opinion convergence through our evolution process. In certain social scenarios, members do expect to acquire efficient agreement e.g. election for superiority. Thus it quite drives our curiosity to investigate how to make opinion gather fast and closely. Specifically, here we are trying to figure out what kind of personality is crucial or necessary in a social activity for consensus.

In Figure 3, (a)-(c) are placement of persons after 200 moments evolution. Compared to Figure 2(c), they have: stronger openness, stronger agreeableness, and weaker neuroticism (higher value for $\sigma$) respectively, as captains present. Obviously openness and agreeableness contribute to opinion convergence given that nodes in Figure 3(a) and Figure 3(b) are denser. This should attribute to reinforced communication. While raising $\sigma$ to a considerable level leads to disorder of entire group. It accords with situations when we meet horror exceptions like disaster.

For leadership, we pick up three lucky person and manipulate their values for leadership. So that their leader identity is assumed. In Figure 4, leaders' trend are expressed as blue arrows. Ideal outcome should be that we authorize several persons, then they summon and encourage others to meet opinion agreement. While as Figure 4(b) presents, though leaders quite gather opinions in a degree, there remains overt communities deviating their track. Associate their initial state (shown in Figure 4(a)) to analyze, the three leaders also gradually changed their opinion trends as the public. In this mean, they are not firm to original idealogy. Thus some followers are lost on the way: they cannot catch up with trend variation all the time. We call such leader as "volatile leader".

Opposite to "volatile leader", leaders could be classified to another genre namely "steady leader". Steady leaders are confident to their opinion and provide reliable support to others. As a result, high convergence is more likely to occur. We model steady leader by assigning no openness to leaders, which makes sense as per above description. Figure 4(c) shows evolution outcomes with three steady leaders. Apparently majority of entities follow their respective leaders. Figure 4(d) is contrast to Figure 4(c). Even if we contrived to inverse leaders' opinion trends at moment 50, similar convergence is observed at moment 200. These comparison sufficient explain the superiority of "steady leader" over "volatile leader".

According to above analysis, we conclude key factors to form a coherent team: members should be highly agreeable and open to diversified ideas, so that sufficient communication can be guaranteed. Safe environment is also crucial, otherwise opinions will not be subjected to any rules. Last but not least, the team requires person with steady view to be leader rather than someone is easily exposed to other perspectives.

Thanks to carefully devised mechanism for network dynamics, we have found some recognized phenomenon in social network by running simulation(opinion convergence\cite{Sobkowicz2011Opinion} and leadership effect\cite{Northouse2014Leadership}). So that our model indeed has promising capacity to verify social science through computation.
\section{prediction}
Apart from modeling network dynamics and interpreting phenomenon in realistic scenes, a desired evolution model should be able to predict personal behaviors. Concretely, personalities and opinions serve as latent variables in our model and can be inferred when network data is observed, then the estimated variables are further used to predict context information in future. In this section, we propose partial training algorithm for PENO model. Further its validity is demonstrated by sufficient experiments.
\begin{figure*}[tt]
    \begin{minipage}{8cm}
            \setlength{\belowcaptionskip}{-0cm}
            \includegraphics[width=8cm,height=6.4cm]{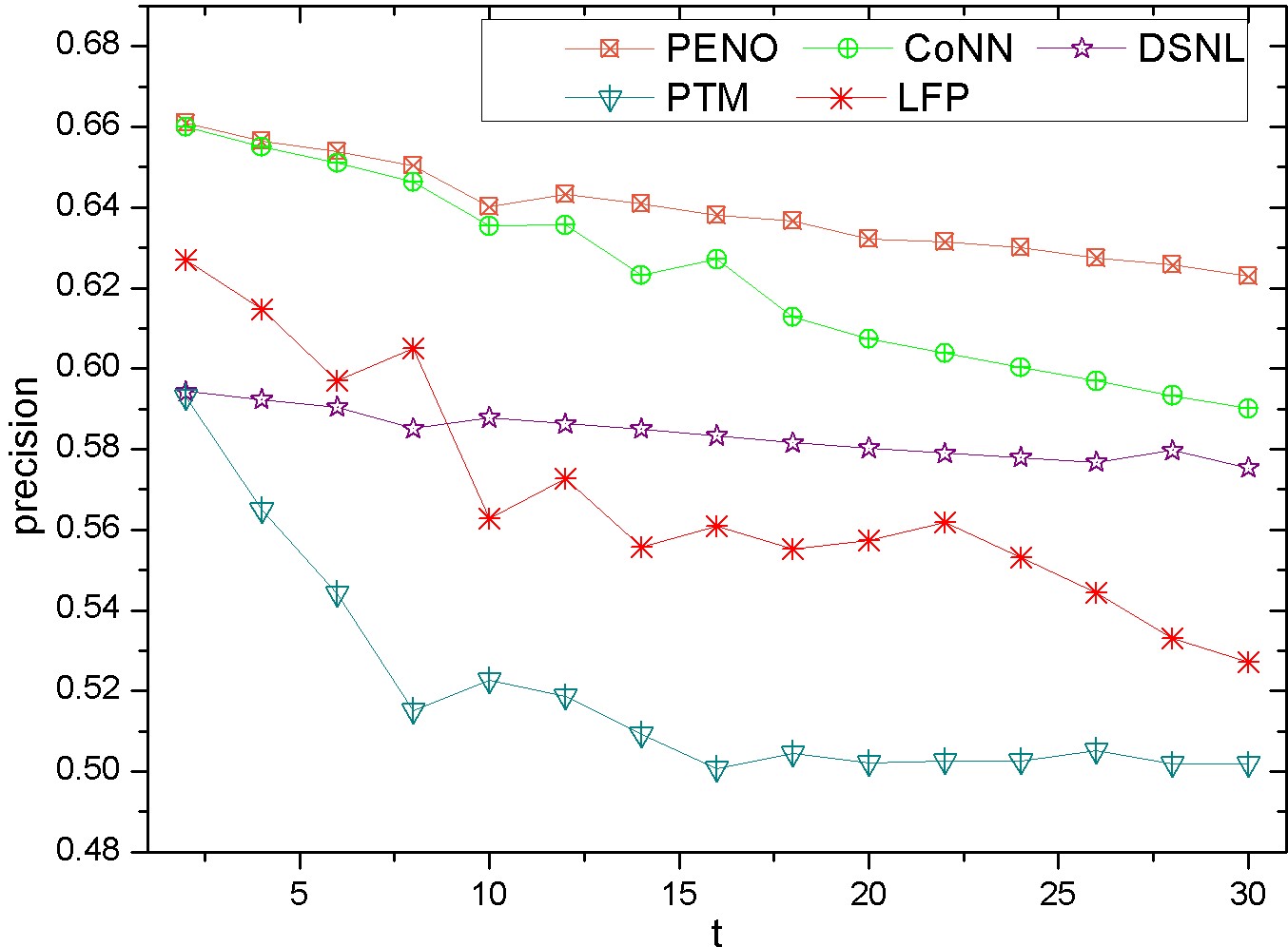}
            \caption*{(a)legislative votes.}
    \end{minipage}
    \begin{minipage}{8cm}
            \setlength{\belowcaptionskip}{-0cm}
            \includegraphics[width=8cm,height=6.4cm]{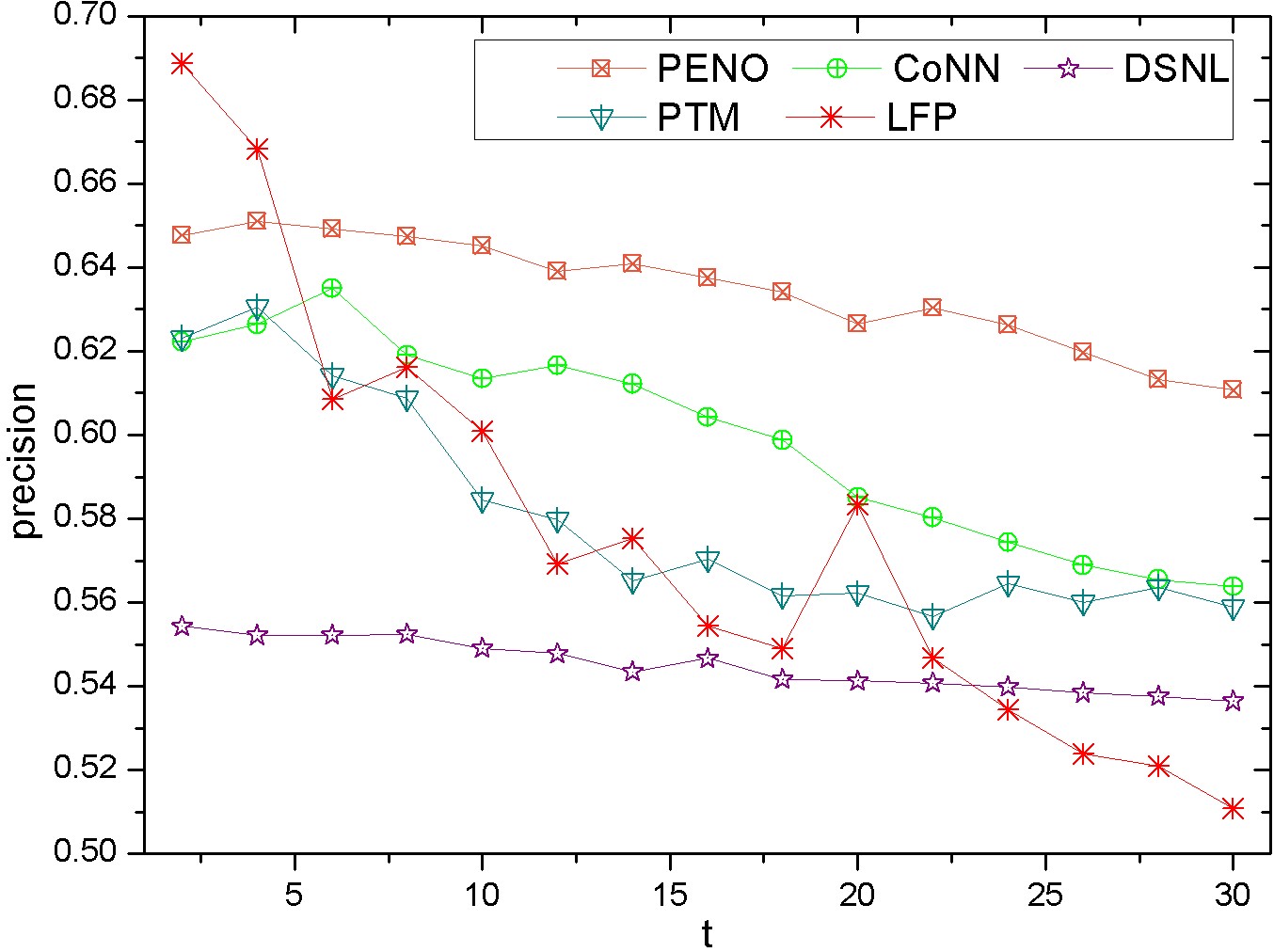}
            \caption*{(b)Digg friendship.}
    \end{minipage}
    \caption{Performances on Predicting for different time span.}
\end{figure*}\\
\subsection{Training}
Since we model both opinion propagation and tie generation into probabilistic process, PENO model could be expressed as a probabilistic graph (shown as Figure 5) and thus have the potential to learn parameters as per Maximum Likelihood Estimation. Our learning objective is to maximize overall probability in the entire evolution process, i.e. joint probability of formed ties and entity properties. Hence parameters should be inferred by
    \begin{equation}
    \setlength{\abovedisplayskip}{3pt}
    \setlength{\belowdisplayskip}{3pt}
    \begin{aligned}
X,\Theta,[\Upsilon,\iota,\varsigma,\beta]=\\ \mathop{argmax}\limits_{X,\Theta,[\Upsilon,\iota,\varsigma,\beta]} & \prod_{t=1}^{T}{P(G^{<t>}|X^{<t>},\theta^{<t>},b)}\\
                                           \times & \prod_{t=2}^{T}P(\theta^{<t>}|\theta^{<t-1>},G^{<t-1>},[\Upsilon,\iota,\beta])
    \end{aligned}
    \end{equation}
where $X = \{x^{<t>}\}_{t=1}^{T}$, $\Theta = \{\theta^{<t>}\}_{t=1}^{T}$,
and $\Upsilon = \{r_{n}\}_{n=1}^{N}$, $\iota =\{l_{n}\}_{n=1}^{N}$,
    $\varsigma =\{\sigma_{n}\}_{n=1}^{N}$,
    $\beta =\{b_{n}\}_{n=1}^{N}$,
    respectively.

Since PENO model is subjected to velocity mechanism(Equation 2-4), it becomes quite arduous for us to derive global optimal solution. Inspired by EM algorithm where latent variable also took important place, we propose an effective approach to train partial parameters in turn. Thanks to continuously refined objective in each turn of optimization, PENO model will eventually be tuned to a state with expected global performance. While one critical issue in training deserves discussion -- step order. Order determines parameters' association and thus optimization gradient. Sometimes it even affects whether the method works well.

Our strategy is to rank parameters as per their association with observed data, i.e. network graph series. Specifically, optimization consists of three steps: (1) train $\beta$; (2)train $[\Upsilon,\iota,\varsigma]$; (3)train X and $\Theta$. In each step, we adopt stochastic gradient ascending method to ensure sampling efficiency.\\
\par \noindent
{\bfseries 5.1.1 train $\beta$}\\ \noindent
As probability graph shows, $\beta$ directly relates to observed data and involves the fewest parameters in inference. Thus training $\beta$ at first becomes spontaneous. Training $\beta$ is equivalent to maximizing joint tie formation
probability:
    \begin{equation}
    \setlength{\abovedisplayskip}{3pt}
    \setlength{\belowdisplayskip}{3pt}
    \beta^{*} = \mathop{argmax}\limits_{\beta}P(F|X, \beta) \\
    \end{equation}
    After stochastic sampling, $\beta$ can be tuned as per the partial derivative of tie generation probability with respect to $\beta$. Given specific tie $e_{ij}^{<t>}$ between person $i$ and $j$ at moment $t$,
    \begin{equation}
    \setlength{\abovedisplayskip}{3pt}
    \setlength{\belowdisplayskip}{3pt}
    \begin{aligned}
                  b_{i}^{*} & = b_i + \frac{\partial log(p_{ij}^{t})}{\partial b_i} \\
                            & = b_i + \frac{-1}{\xi^{2}}\times\frac{||x_i^t-x_j^t||}{-b_i^2*b_j}
    \end{aligned}
    \end{equation}
$b_{j}^{*}$ is derived in the same way.\\
\par \noindent
{\bfseries 5.1.2 train $X$ and $\Theta$}\\ \noindent
After tuning $\beta$, rest relation in inference is: $[\Upsilon,\iota,\varsigma] \rightarrow \Theta \rightarrow X $. Once $\Theta$ is estimated, $X$ will also be fixed as per velocity. Hence then we should train $\Theta$ and $X$.
\\
As $\Theta$ and $X$ are correlated with respect to adjacent moments, they are also tuned in a time-dependent pattern. At moment 0, opinion can be inferred by maximizing the probability of initial observed relation graph:
    \begin{equation}
    \setlength{\abovedisplayskip}{3pt}
    \setlength{\belowdisplayskip}{3pt}
    X^{<0>*} = \mathop{argmax}\limits_{X^{<0>}}log(P(G^{<0>}|X^{<0>}, \beta)) \\
    \end{equation}

Given a stochastic sampled tie $e_{ij}^{<0>}$ between person $i$ and $j$ at initial moment $0$, $x_i^{<0>} = <m_i^{<0>},n_i^{<0>}>$ is estimated as per tie generation rule:
    \begin{equation}
    \setlength{\abovedisplayskip}{3pt}
    \setlength{\belowdisplayskip}{3pt}
    \begin{aligned}
                   z_i^{<0>*} & = z_i^{<0>} + \frac{\partial log(p_{ij}^{<0>})}{\partial z_i^{<0>}} \\
                            & = z_i^{<0>} + \frac{-1}{\xi^{2}}\times \frac{2(z_i^{<0>} - z_j^{<0>})}{b_i * b_j}
    \end{aligned}
    \end{equation}
    $z_j^{<0>*},c_i^{<0>*},c_j^{<0>*}$ are estimated as the same way. Then we are able to derive $x^{<1>*}$ based on Equation (3) and (4), $x^{<0>*}$, and $\theta^{<0>}$.
    After that, at each moment $t (s.t. \  t>1)$, as long as $\theta^{<t-1>}, G^{<t-1>}$ and $X^{t}$ are provided, $\theta^{<t>}$ is estimated by:
    \begin{equation}
    \setlength{\abovedisplayskip}{3pt}
    \setlength{\belowdisplayskip}{3pt}
    \begin{aligned}
     \theta^{<t>*} = \mathop{argmax}\limits_{\theta^{<t>}}[log
                   & P(\theta^{<t>}|\theta^{<t-1>},G^{<t-1>},[\Upsilon,\iota,\varsigma]) \\ \times
                   & P(G^{<t+1>}|\theta^{<t>},X^{<t>},\beta)] \\
    \end{aligned}
    \end{equation}
    Given specific tie $e_{ij}^{<t>}$ between person $i$ and $j$ at moment $t$,
    \begin{equation}
    \setlength{\abovedisplayskip}{3pt}
    \setlength{\belowdisplayskip}{3pt}
    \begin{aligned}
    \theta_i^{<t>*}
            & = \theta_i^{<t>} + \frac{\partial log(p_{ij}^{<t+1>})}{\partial \theta_i^{<t>}} \\
            & + \frac{\partial logN(\theta_i^{<t-1>} + r_{i}\mathop{\sum}\limits_{(i,k) \in G^{<t-1>}
                 }\frac{l_{k}}{\lambda_i^{<t-1>}}\triangle\theta_{ki}^{<t-1>},\sigma_i^2)}{\partial \theta_i^{<t>}} \\
            & = \theta_i^{<t>} + \frac{\partial log(p_{ij}^{<t+1>})}{\partial \theta_i^{<t>}} \\
            & + \frac{-\theta_i^{<t>} + \theta_i^{<t-1>} + r_i\mathop{\sum}\limits_{(i,k) \in G^{<t-1>}
                 }\frac{l_k}{\lambda_i^{<t-1>}}\triangle\theta_{ki}^{<t-1>} }{\sigma_i^2}\\
    \end{aligned}
    \end{equation}
    \begin{equation}
    \setlength{\abovedisplayskip}{3pt}
    \setlength{\belowdisplayskip}{3pt}
    \begin{aligned}
    \frac{\partial log(p_{ij}^{<t+1>})}{\partial \theta_i^{<t>}}
            & = \frac{-1}{\xi^{2}b_{i}b_{j}} \frac {\partial ||x_i^{<t+1>} - x_j^{<t+1>}||^2}{\partial
                \theta_i^{<t>} } \\
            & = \frac{-1}{\xi^{2}b_{i}b_{j}} \frac {\partial (m_i^{<t+1>} - m_j^{<t+1>})^2}
                {\partial \theta_i^{<t>}} \\
            & + \frac{-1}{\xi^{2}b_{i}b_{j}} \frac {\partial (n_i^{<t+1>} - n_j^{<t+1>})^2}
                {\partial \theta_i^{<t>}}
    \end{aligned}
    \end{equation}

\begin{figure}[tt]
    \begin{minipage}{4cm}
        \setlength{\belowcaptionskip}{-0cm}
        \includegraphics[width=4cm]{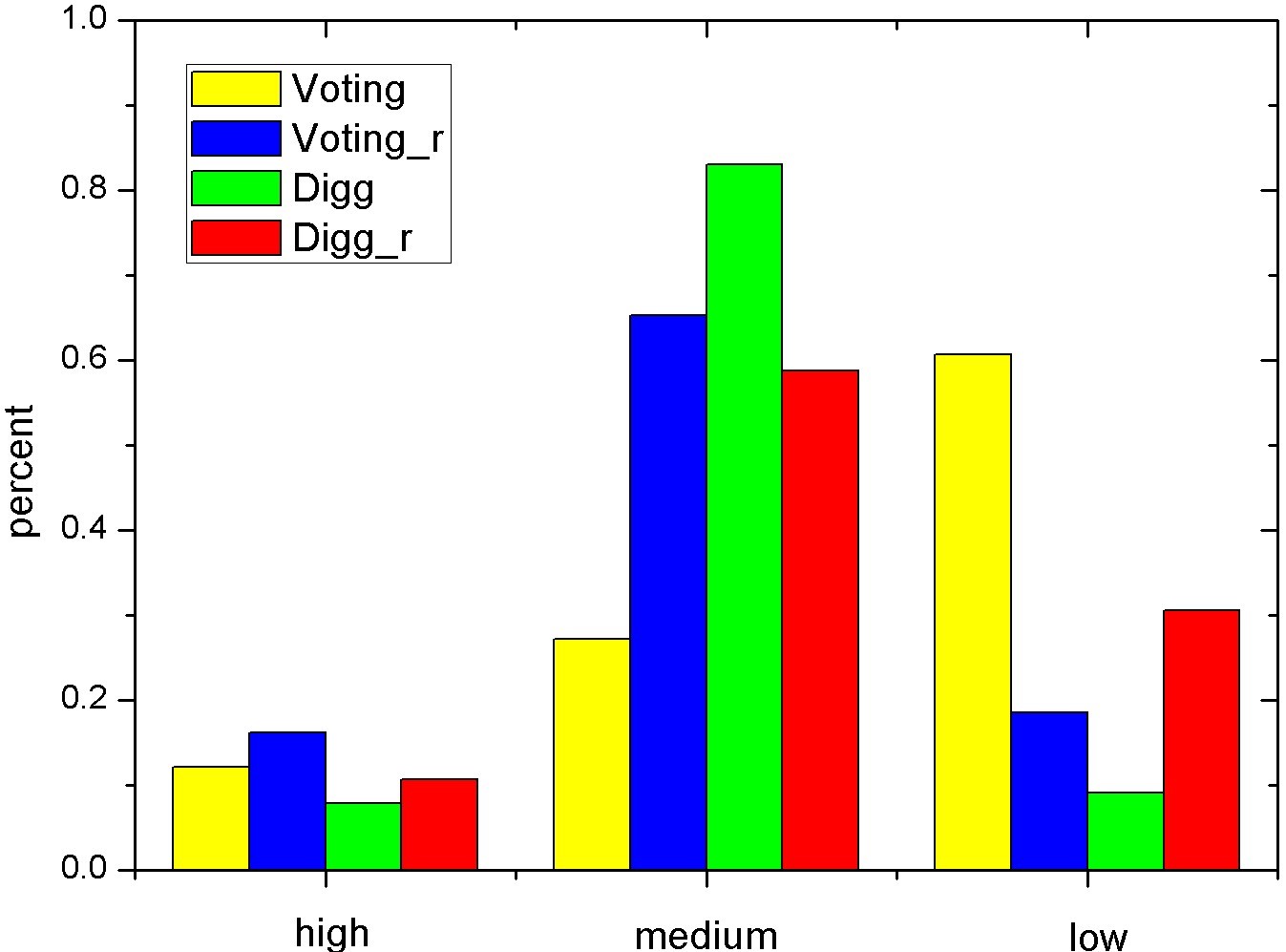}
        \caption*{(a)agreeableness.}
    \end{minipage}
    \begin{minipage}{4cm}
        \setlength{\belowcaptionskip}{-0cm}
        \includegraphics[width=4cm]{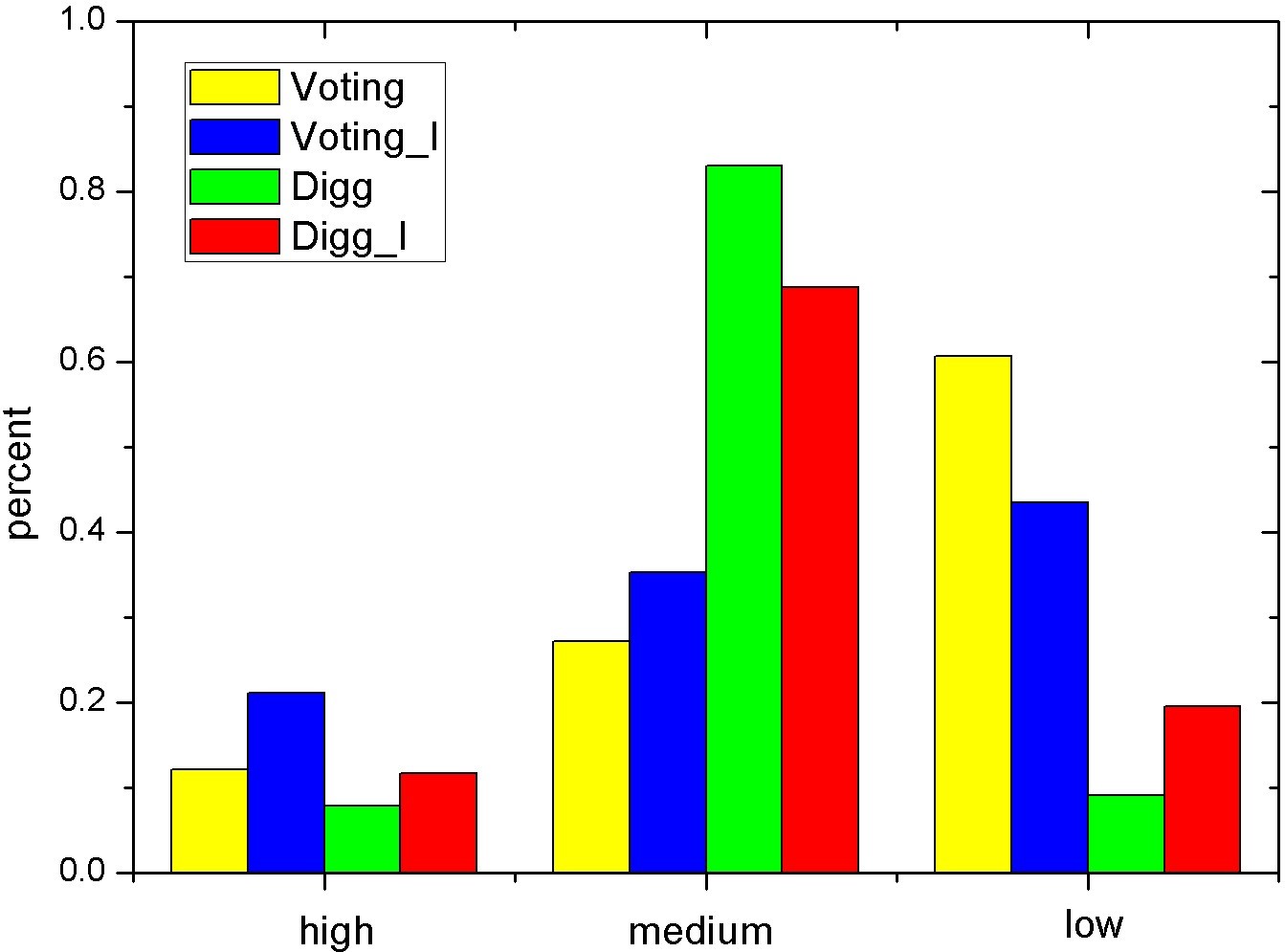}
        \caption*{(b)leadership.}
    \end{minipage}

    \begin{minipage}{4cm}
        \setlength{\belowcaptionskip}{-0cm}
        \includegraphics[width=4cm]{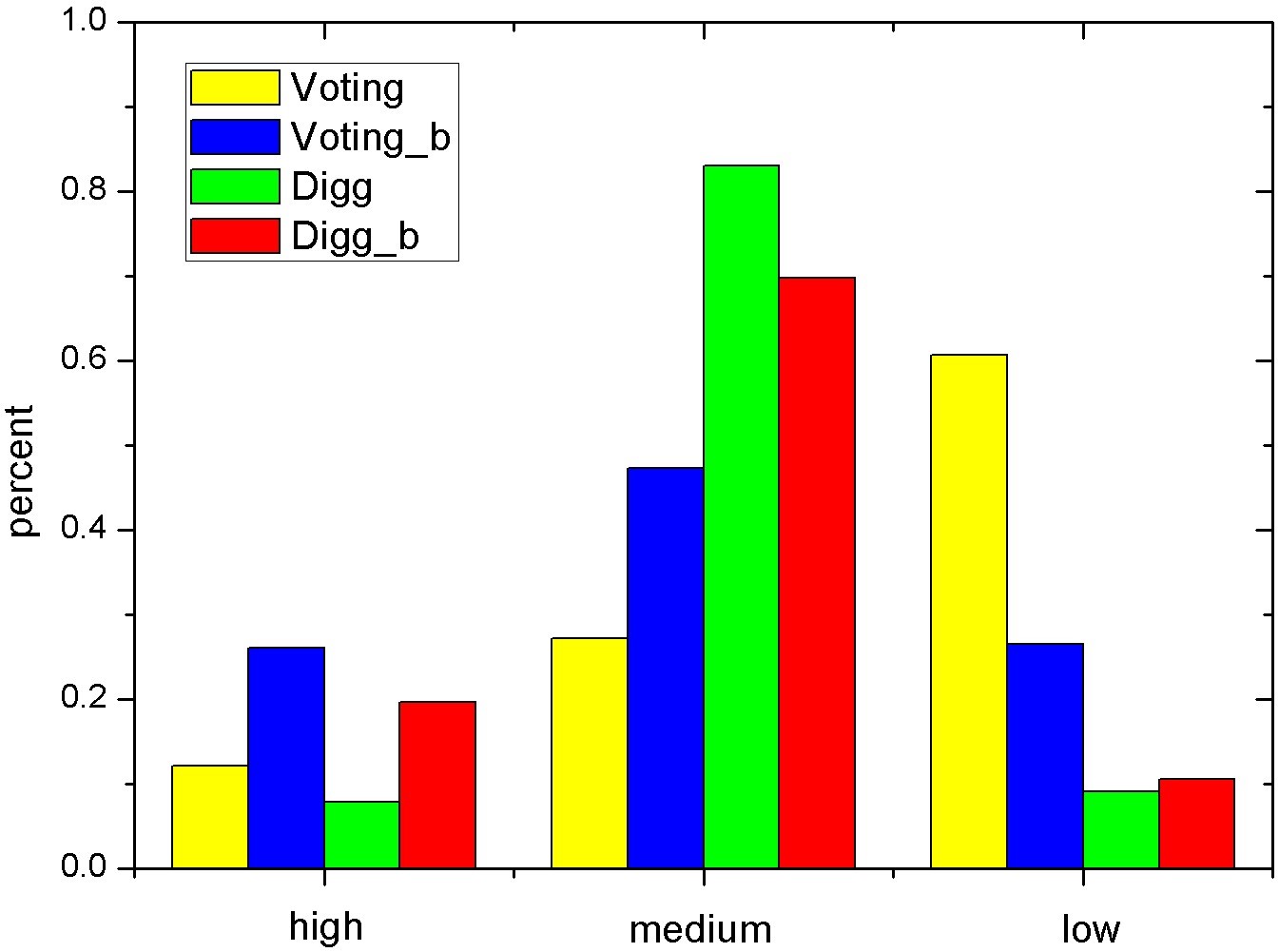}
        \caption*{(c)openness.}
    \end{minipage}
    \begin{minipage}{4cm}
        \setlength{\belowcaptionskip}{-0cm}
        \includegraphics[width=4cm]{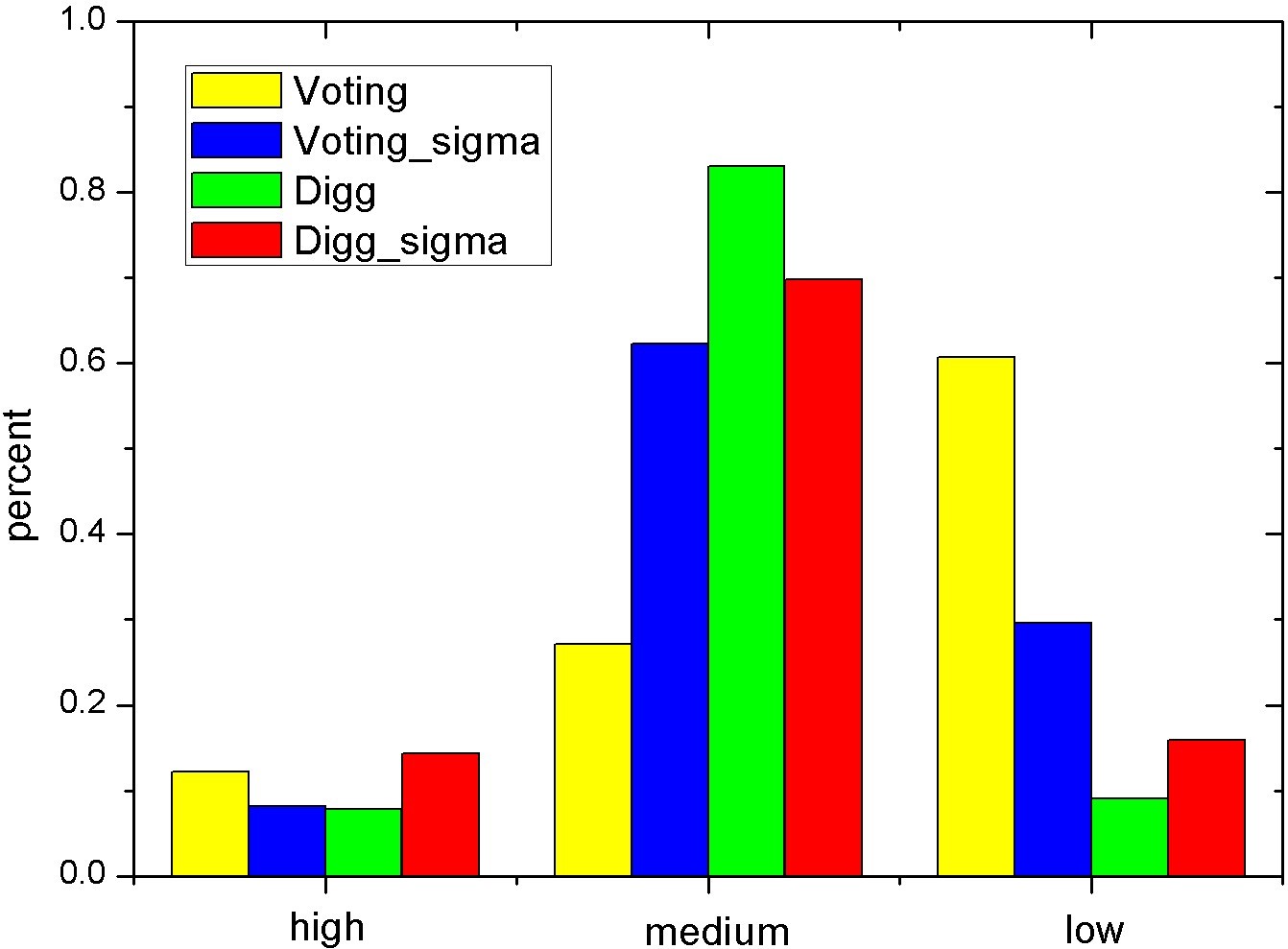}
        \caption*{(d)neuroticism.}
    \end{minipage}
    \caption {Personalities Distribution in different groups.}
\end{figure}
According to Equation 7, we can infer
    \begin{equation}
    \setlength{\abovedisplayskip}{3pt}
    \setlength{\belowdisplayskip}{3pt}
    \begin{aligned}
    \frac{\partial log(p_{ij}^{<t+1>})}{\partial \theta_i^{<t>}}
            & = \frac{-2v}{\xi^{2}b_{i}b_{j}} [(m_i^{<t+1>} - m_j^{<t+1>})\cdot (-sin\theta_i^{<t>}) \\
            & \ \ \ \ \ \ \ \ \
              + (n_i^{<t+1>} - n_j^{<t+1>})\cdot cos\theta_i^{<t>}]
    \end{aligned}
    \end{equation}
\par \noindent
{\bfseries 5.1.3 train $[\Upsilon,\iota,\varsigma]$}\\ \noindent
Similar to $\Theta$ and $X$, $\Upsilon$, $\iota$, and $\varsigma$ contribute to opinion dynamics as well. It is nice to tune them
according to time series. Objective in this step is
    \begin{equation}
    \setlength{\abovedisplayskip}{3pt}
    \setlength{\belowdisplayskip}{3pt}
    \begin{aligned}
    \\ [R^*,L^*,\sigma^*] = \mathop{argmax}\limits_{[R,L,\sigma]}P([R,L,\sigma]|F,\theta)
    \end{aligned}
    \end{equation}
    Given specific moment $t$, $\Upsilon$,$\iota$, and $\varsigma$ are respectively tuned as per the partial derivative of opinion evolution probability with respect to them at moment $t$:\\
    \begin{equation}
    \setlength{\abovedisplayskip}{3pt}
    \setlength{\belowdisplayskip}{3pt}
    \begin{split}
    r_i^* = r_i & + \frac{\partial logN(\theta_i^{<t-1>} + \mathop{\sum}\limits_{(i,k) \in G^{<t-1>}
                 }\frac{r_{i}l_{k}}{\lambda_i^{<t-1>}}\triangle\theta_{ki}^{<t-1>},\sigma_i^2)}{\partial r_i} \\
          = r_i & + (\theta_i^{<t>}-\theta_i^{<t-1>}-\mathop{\sum}\limits_{(i,k) \in G^{<t-1>}
                 }\frac{r_{i}l_{k}}{\lambda_i^{<t-1>}}\triangle\theta_{ki}^{<t-1>})  \\
                & \times\frac{\mathop{\sum}\limits_{(i,k) \in G^{<t-1>}}\frac{l_k}{\lambda_i^{<t-1>}}\triangle\theta_{ki}^{<t-1>}
                 }{\sigma_i^2}\\
    \end{split}
    \end{equation}
    \begin{equation}\nonumber
    \setlength{\abovedisplayskip}{3pt}
    \setlength{\belowdisplayskip}{3pt}
    \sigma_i^* = \sigma_i + \frac{\partial logN(\theta_i^{<t-1>} + \mathop{\sum}\limits_{(i,k) \in G^{<t-1>}
                 }\frac{r_{i}l_{k}}{\lambda_i^{<t-1>}}\triangle\theta_{ki}^{<t-1>},\sigma_i^2)}{\partial \sigma_i} \\
    \end{equation}
    \begin{equation}
    \setlength{\abovedisplayskip}{3pt}
    \setlength{\belowdisplayskip}{3pt}
               = \frac{\sigma_i^4-\sigma_i^2 + \theta_i^{<t>}-\theta_i^{<t-1>}-\mathop{\sum}\limits_{(i,k) \in G^{<t-1>}
                 }\frac{r_{i}l_{k}}{\lambda_i^{<t-1>}}\triangle\theta_{ki}^{<t-1>}}{\sigma_i^3}\\
    \end{equation}
Given a stochastic sampled tie $e_{ij}^{<t>}$ between person $i$ and $j$ at initial moment $t$,
    \begin{equation*}
    \setlength{\abovedisplayskip}{3pt}
    \setlength{\belowdisplayskip}{3pt}
    \begin{split}
    l_i^* = l_i & + \frac{\partial logN(\theta_j^{<t-1>} + \mathop{\sum}\limits_{(j,k) \in G^{<t-1>}
                 }\frac{r_{j}l_{k}}{\lambda_j^{<t-1>}}\triangle\theta_{kj}^{<t-1>},\sigma_j^2)}{\partial l_i} \\
                = l_i & + [(\theta_j^{<t>}-\theta_j^{<t-1>}-\mathop{\sum}\limits_{(j,k) \in G^{<t-1>}
                 }\frac{r_{j}l_{k}}{\lambda_j^{<t-1>}}\triangle\theta_{kj}^{<t-1>})]\\
    \end{split}
    \end{equation*}
        \begin{equation}
    \setlength{\abovedisplayskip}{3pt}
    \setlength{\belowdisplayskip}{3pt}
    \begin{split}
                & \times \frac{r_j(\lambda_j^{<t-1>}-l_i)}{(\lambda_j^{<t-1>})^2\sigma_j^2}\triangle\theta_{ji}^{<t-1>}\\
    \end{split}
    \end{equation}
So far, all parameters have been tuned in the training turn. Such strategy possesses another overt advantage: each parameter is optimized later than its predecessor in inference. Thus they would not interfere mutually when optimization process is ongoing.
\subsection{Data}
In experiments, we base model training on two typical datasets with respect to idealogy:

\textit{{\bfseries Legislative voting.}} \noindent
In congress, a new bill should be introduced by a sponsor for consideration. After that, other members vote for it to present support or not one by one. Voting records convey direct information about political stand and social ties. Collected data contain bills from 1983 to 2017. Minimum moment interval is set to one month, and we get voting records within one year together as network snapshot to smooth evolution process. Entire evolution contains $T = 382$ moments, $N = 2,180$ legislators, $130,692$ bills and 2.1 million voting records.

\textit{{\bfseries Digg friendship.}} \noindent
Digg is a news website where users recommend stories and articles to the public. Friendship on Digg exactly indicates the users sharing similar idealogy. Minimum moment interval is set to three days. We assume that friendship will not spontaneously break. In total, $T = 209$ moments, selected users $N = 5,000$, and 71,985 friendship ties are considered.
\subsection{Experiment}

{\bfseries Performance comparison.} \noindent
To demonstrate the effectiveness of our model, we compare its performance on prediction tasks with several the-state-of-arts dynamic network baselines: \noindent
\begin{itemize}[leftmargin=8pt]
\item{Co-evolve social network model(CoNN)\cite{DBLP_conf_kdd_GuSG17}: The basic model where network structure and latent feature
affect reciprocally and evolve at the same time. The authors kindly share their code and data.}
\item{Dynamic social network with latent space models (DSNL)\cite{Sarkar2005Dynamic}: This model updates properties as per distance-based
    probability. For fair comparison, we remove its trick on initialization.}
\item{Latent feature propagation model(LFP)\cite{Heaukulani2013Dynamic}: Primary idea relies on binary latent feature propagation. It aims
to mine connection from finite common behaviors or interests.}
\item{Phase transition model(PTM)\cite{Benjacob1995Novel}: An approach to describe molecule behaviors during phase transition. We set its
properties equal to ours when the first moment requiring prediction comes.}
\end{itemize}
Dimension of latent variable is fixed at 2, and other settings are included in Table 1. Later analysis reveals that system parameters have no severe impact on our model within certain ranges.

\begin{table}[h]
    \setlength{\abovecaptionskip}{0.1cm}
    \center \caption{System Settings}

    \begin{tabular*}{6cm}{c c c}
        \toprule
        {\bfseries parameter} & \ \ \ \ \ \ & {\bfseries value}\\
        \midrule
        {\bfseries velocity} & \ \ \ \ \ \ & $3\times10^{-3}$\\
        {\bfseries iteration times} & \ \ \ \ \ \ & 5\\
        {\bfseries $\xi$} & \ \ \ \ \ \ & {0.6} \\
      \bottomrule
    \end{tabular*}
\end{table}

\begin{figure}
    \includegraphics[width=9cm]{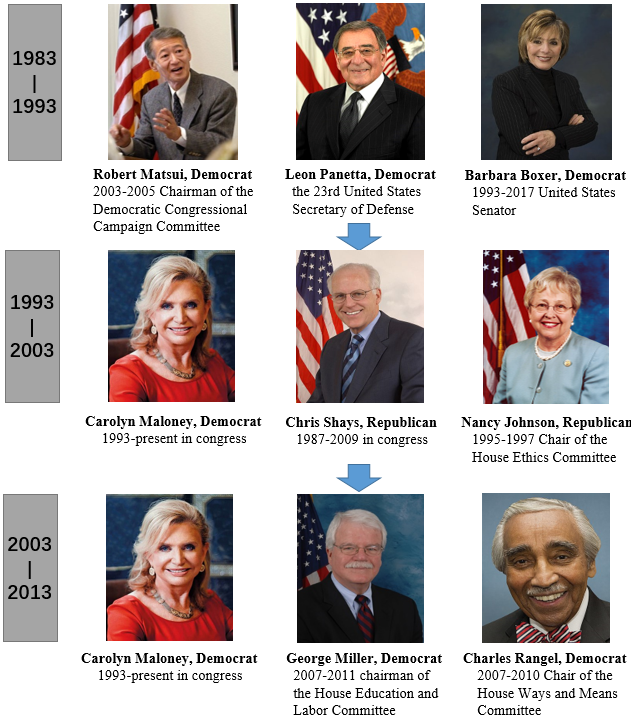}
    \caption{Estimated leaders in congress over time.}
\end{figure}

Figure 6 shows prediction performance given different time span to predict. The span $t$ indicates that if network graph series from initial moment $t_0$ to a certain moment $t^*$ are observed, our goal is to estimate network evolution till the last moment $t^*+t$. We rank each tie as per its formation probability and then evaluate AUC scores, averaged on all time span ($t$,$t^*+t$] within data range.

\textit{{\bfseries Personality distribution.}} \noindent
As PENO incorporates personality effect, it would also be interesting to study what it is alike through learning. Figure 7 presents distribution of the four personality genres in groups, respectively. For each genre, we classify a person according to its value:(1) higher than average by more than $20\% \rightarrow$ "high"; (2) lower than average by more than $20\% \rightarrow$ "low";(3) otherwise $\rightarrow$ "medium". And columns for "Voting" and "Digg" in subgraphs are presented as reference. Their classification is based on ties of the person have in raw data.

We can find that voting group has more percents of persons with both high leadership and low leadership than Digg. Its overall agreeableness average also exceeds. This should attribute to their difference in system structure: Congress is highly Centralized by parties and leaders, while Digg encourage everyone to share opinions. Neuroticism comparison also reveals that voting activity is much more intentional than article recommendation.

\textit{{\bfseries Times leader.}} \noindent
In voting prediction, we also extract the congressmen with top 3 high leadership for each decade. This provides us another perspective to take a glance on times political leaders. As Figure 8 presents, we indeed discover some names who are highly motivated to promote legislative bills like Robert Matsui and Carolyn Maloney \cite{Carolyn_Maloney, Izenstark2014Asian}. And Leon Panetta is the former United States Secretary of Defense. It is noteworthy that they three are exactly well-known for firm political stand, which accords to our findings in simulation section.

\textit{{\bfseries Sensitivity Analysis.}} \noindent
In figure 9, model performance are shown with different settings for system parameters (time span $t$ are fixed at 10). Obviously, they will not cause decisive impact on learning evolution.

\begin{figure}[h]
    \begin{minipage}{3.75cm}
        \setlength{\belowcaptionskip}{-0cm}
        \includegraphics[width=3.75cm]{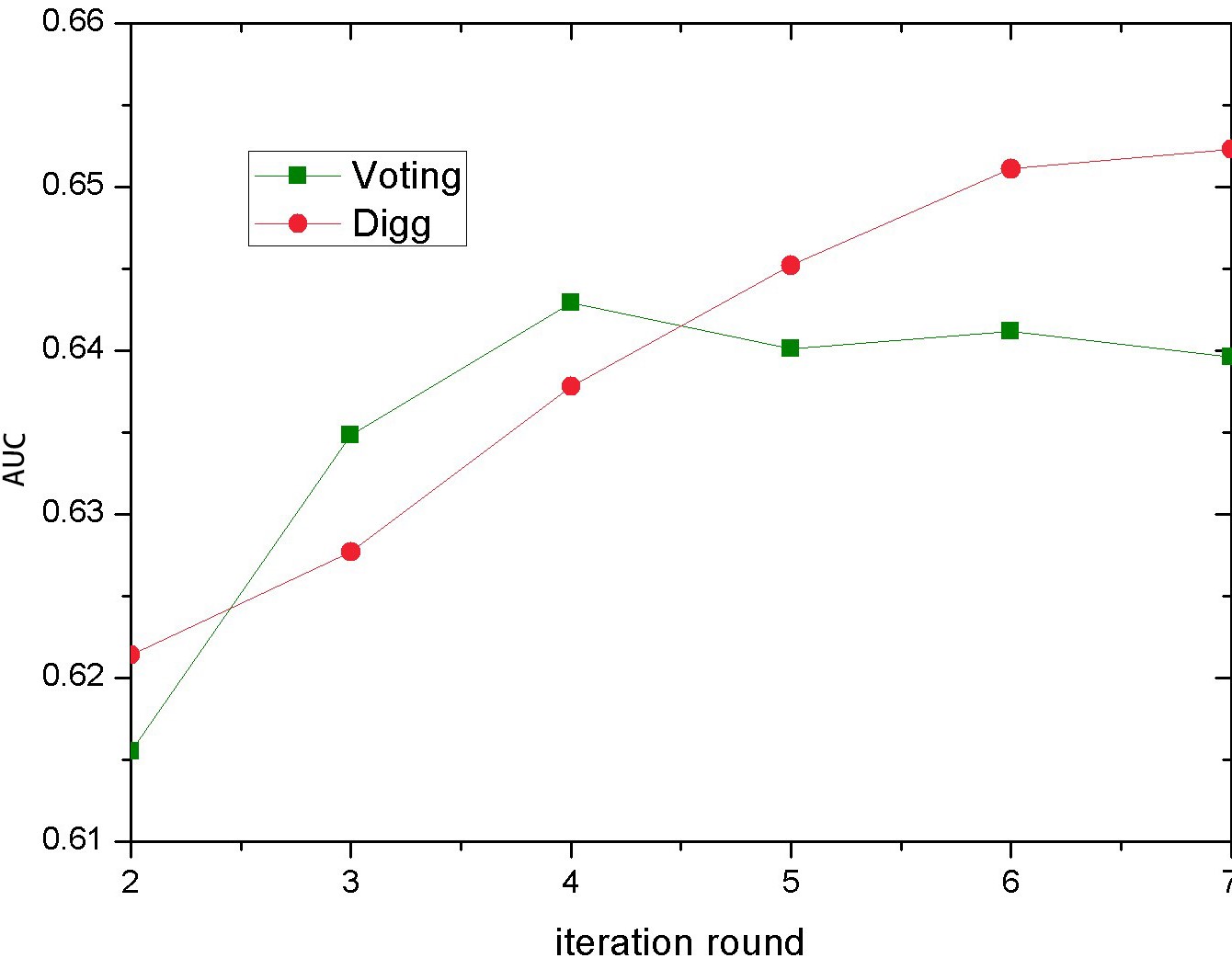}
        \caption*{(a) iteration rounds.}
    \end{minipage}
    \ \ \ \
    \begin{minipage}{3.85cm}
        \setlength{\belowcaptionskip}{-0cm}
        \includegraphics[width=3.85cm]{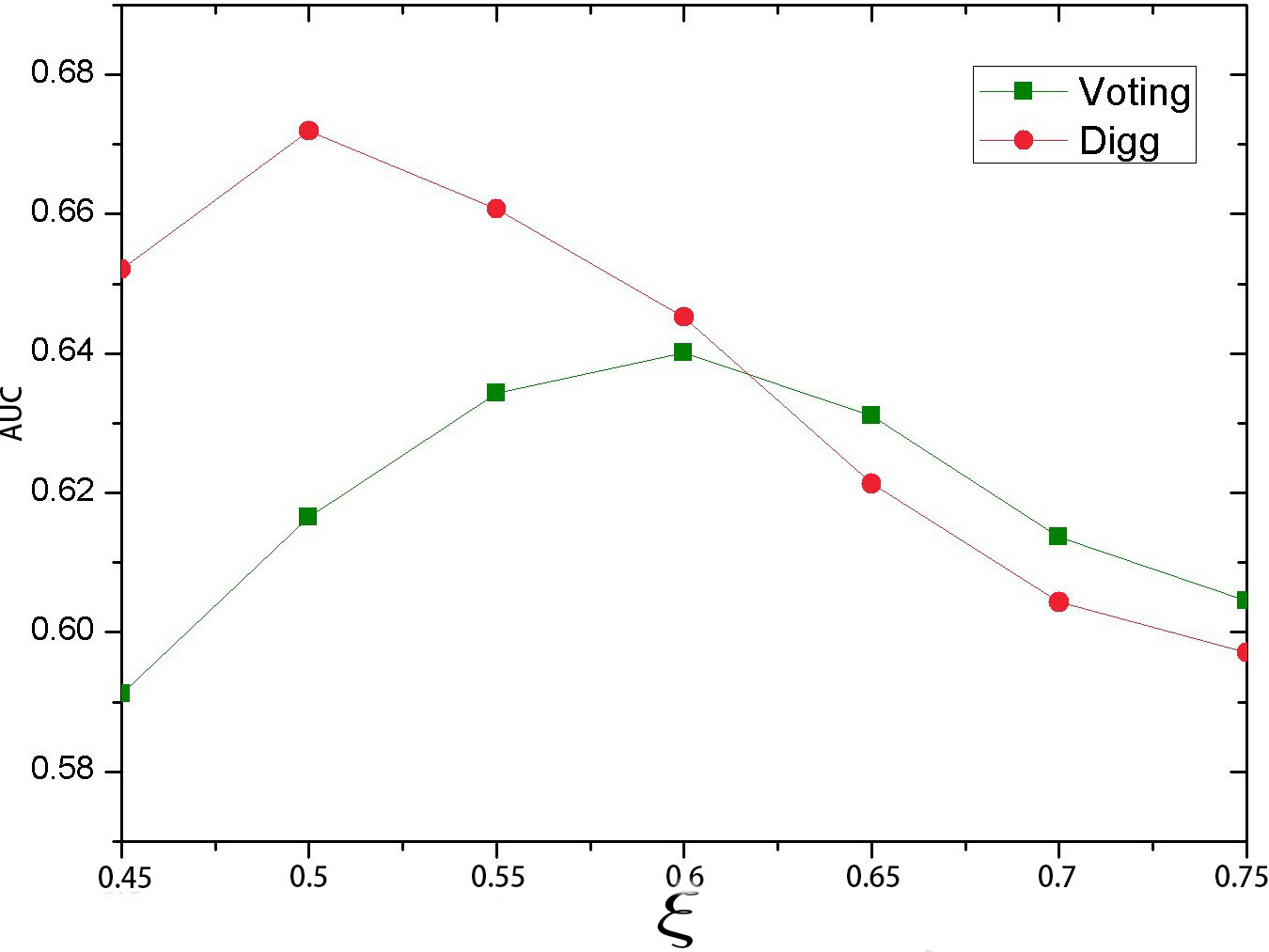}
        \caption*{(b) $\xi$.}
    \end{minipage}
    \caption {Parameter study on system parameters.}
\end{figure}

\section{Discussion}
Unlike previous approaches to describe dynamic network, PENO introduces personality factors into co-evolution process of entity properties and ties. By learning parameters, our model has potential to profile a group of users given time series of their social relationship, and further extends to a variety of human-centered application - recommending article style, predicting travel route, or arranging exercise plan. These tasks always involves both social relationship and personal preference. So if PENO could be combined to other methods that are focused on dealing with private information in concrete scenes, there will be a great chance to learn multi-tasks at the same time, or to transfer experience of one type to another by sharing latent features or inheriting property values as initialization. Figure 10 give an example to illustrate our vision.

\section{Related work}
\textit{{\bfseries Dynamic Network.}} \noindent
Understanding how social network evolves with links and properties has been a research hotspot recently. The most classical type of network model is latent space model\cite{Hoff2002Latent, Xing2010A, Fu2009Dynamic}. It assumes the snapshot of a social
network is generated according to the positions of individuals in an unobserved social space. While the position usually play a implicit role and cannot convey intuitive meaning in social science as they are latent. Another common idea is based on hidden Markov model, where latent variables merely depends on their last values\cite{Corneli2015Modelling, Heaukulani2013Dynamic, Zheleva2009Co}. And evolution of latent features is regarded as regression of future variable series of the entities to accommodate dynamic networks. These methods, however, fail to consider the mutual impact between variable and network structure during evolution\cite{DBLP_conf_kdd_GuSG17}. Also, none of aforementioned models attempts to understand evolving network from mental analysis at individual level.

\begin{figure}[t]
    \setlength{\abovecaptionskip}{0.1cm}
    \setlength{\belowcaptionskip}{-0.1cm}
    \includegraphics[width=9cm]{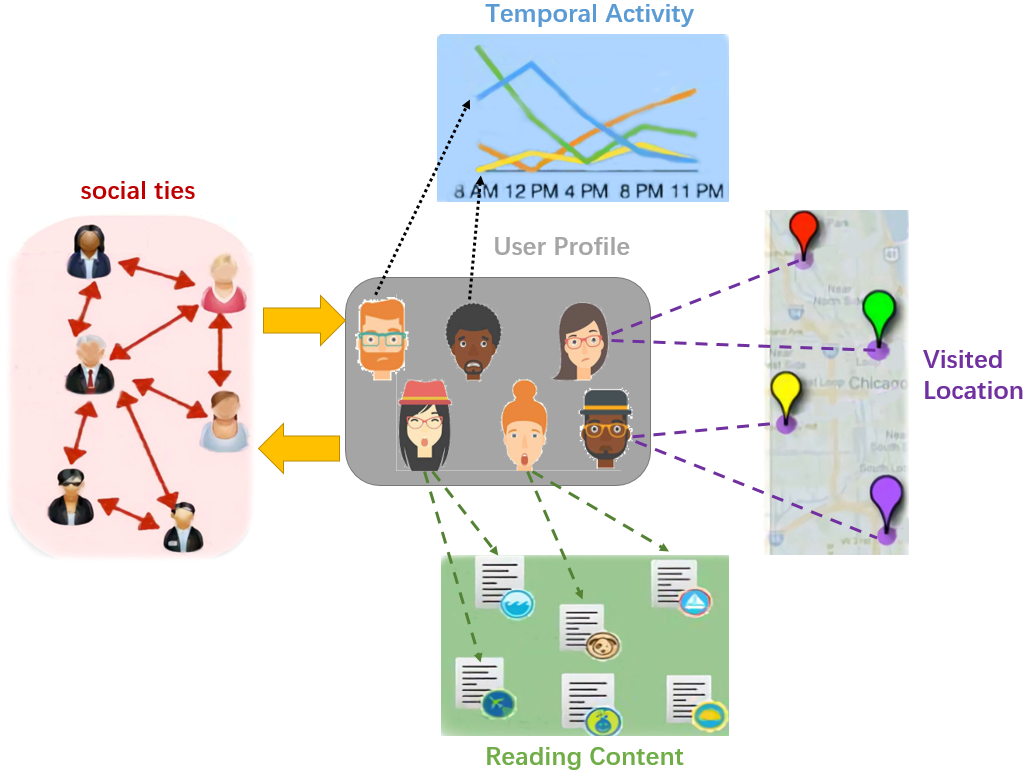}
    \caption{The vision of PENO in multi-task learning.}
\end{figure}

\textit{{\bfseries Social Psychology.}} \noindent
Other than computer model, psychology study investigates network as group behavior. Individual spontaneously presents innate characteristic, which usually attributes to personality, and group feature formed by social influence \cite{Krull2001Multilevel, Tang2009Social, Tang2013Confluence}. As per Big five personality classification\cite{MURRAY2010THE}, the agreeableness of a person indicates his willing to accept views from others. Openness stands for whether he would like to make friends with others, and neuroticism reveals his emotional stability, even this is easily influenced by environment factor. Inverse to agreeableness, another genre namely leadership is proposed for revealing someone's ability in persuading people\cite{Northouse2014Leadership}. In information network, idealogy or opinion exactly suits to represent the person so that opinion propagation becomes a prevalent form of social influence \cite{Acemoglu2011Opinion, Liu2006Public}. Although all these illustration seems reasonable, these studies have to face huge difficulty in collecting data from questionnaires. Hence, their conclusion are unable to be supported by large-scale data.
\section{Conclusion}
In this work, we propose PENO, a dynamic network model for understanding interaction between social relationship, individual idealogy, and innate personality. Everyone produces his unique impact to friends and meanwhile receive reciprocal impact. So that idealogy propagates via ties over time. Then idealogy affinity will also be reappraised to grasp a snapshot of network graph. Simulation demonstrates the contribution of personality to group idealogy and further concludes prerequisites of achieving agreement. It is a promising method to verify social science by computation. In prediction tasks on two typical dataset(legislative votes and Digg friendship), experiments exhibit not only our advantage over the state-of-the-arts, but also the capacity to reveal social feature of a group and its leaders during different periods. PENO gives the very first attempt to understand network evolution from a social psychology perspective. In future, we expect to fuse it with other approaches in more human-centered applications by multi-task learning and transfer learning.

\bibliographystyle{ACM-Reference-Format}
\bibliography{sample-bibliography}

\end{document}